%% file: main.tex
\documentclass[11pt]{article}

\usepackage[letterpaper,margin=1.25in]{geometry}
\usepackage{amsmath}   
\usepackage{amssymb}
\usepackage{graphicx}
\usepackage{tikz}
\usepackage{bm}
\usepackage{newtxtext}
\usepackage{newtxmath}
\usepackage{natbib}
\usepackage{hyperref}
\usetikzlibrary{arrows.meta,positioning,backgrounds,fit,decorations.pathreplacing,calc}

\newcommand{\aff}[1]{\textsuperscript{#1}}
\newcommand{\backsection}[2][]{\medskip\noindent\textbf{#1.}~#2\par}

\title{A source-term interpretation of turbulent rough wall-pressure spectra}
\author{%
J. M. O. Massey\textsuperscript{1}\thanks{Email: \texttt{masseyj@stanford.edu}},
A. J. Smits\textsuperscript{2}
and B. J. McKeon\textsuperscript{1}}
\date{%
\aff{1}Center for Turbulence Research, Stanford University, Stanford, CA 94305, USA\\
\aff{2}Department of Mechanical and Aerospace Engineering, Princeton University, Princeton, NJ 08544, USA}

\newcommand{\ks}{k_s}
\newcommand{\dd}{\mathrm{d}}
\newcommand{\kvec}{\bm{k}}
\newcommand{\ue}{U_e}
\newcommand{\Lin}{\mathcal{L}}
\newcommand{\NL}{\mathcal{Q}}

\begin{document}

\maketitle

\begin{abstract}
Wall-pressure fluctuations beneath a turbulent boundary layer drive the flow-induced noise and structural loading of rough surfaces.  On a smooth wall their variance grows with the logarithm of the Reynolds number, a growth carried, in a source-term reading, by the nonlinear turbulence--turbulence part of the pressure source acting across the logarithmic layer.  We ask what sets the same growth once the wall is fully rough.  Standard rough-wall phenomenology answers it: above the roughness the mean flow keeps its smooth-wall logarithmic form, so the active source range is cut off at the roughness height rather than the viscous length, and the variance grows on the logarithmic span between the roughness height and the layer thickness in place of the Reynolds number.  Splitting the pressure source into its two physical parts then distinguishes separate contributions to the energy: a roughness-local one from the mean-shear source, fixed at high frequency on the roughness scale, and an energetic one from the nonlinear source at the outer scale, which alone carries the growth.  Calibrated against rough-wall cases, the model collapses the spectral shape onto this energetic peak and holds it Reynolds-independent at fixed geometry.  The canopy contribution decays with distance from the wall and contributes a finite offset. The growth coefficient is predicted to be the smooth-wall one.  The present range is too narrow to discriminate that rate from twice or half it, so the contribution is a parameter-free prediction and the identification of a suitable roughness-size sweep that would test it.
\end{abstract}

\section{Introduction}\label{sec:intro}
Wall-pressure fluctuations beneath a turbulent boundary layer drive flow-induced noise and structural loading, and their frequency spectrum is the central input to rough-wall noise prediction \citep{blake1970,goody2004}.  On a smooth wall the inner-scaled integrated variance grows with Reynolds number \citep{farabee1991,bull1996}, and can be represented logarithmically by $\overline{p^{\prime 2}}^+ = A\ln\delta^+ + \mathrm{const}$, where $\delta^+\equiv\delta u_\tau/\nu$ is the friction Reynolds number built from the boundary-layer thickness $\delta$, the friction velocity $u_\tau$ and the kinematic viscosity $\nu$, and superscript $+$ denotes viscous scaling throughout, with $A\approx 2.3$ from fits to the data of \citet{schlatterorlu2010,leemoser2015} (see \S\ref{sec:model}) and the analyses of \citet{panton2017,massey_eddy_2025}.  This growth is the wall imprint of pressure sources distributed across the logarithmic layer, whose active range widens as $\ln\delta^+$.  In a source-term reading the fluctuating kinematic pressure $\pi'$ satisfies the Poisson equation
\begin{equation}
    \partial_i\partial_i\pi' = \Lin+\NL,
    \qquad
    \Lin=-2(\partial_j\bar U_i)(\partial_i u'_j),
    \qquad
    \NL=-\partial_i\partial_j\!\left(u'_iu'_j-\overline{u'_iu'_j}\right),
    \label{eq:poisson}
\end{equation}
with $\bar U_i$ the mean velocity, $u'_i$ the velocity fluctuation, $\partial_i\equiv\partial/\partial x_i$ and an overbar the time average: the variance separates into a linear source $\Lin$ (the product of mean shear and fluctuating strain; the `rapid' source of the pressure-strain literature) and a nonlinear source $\NL$ quadratic in the fluctuations (`slow') \citep{kim1989,changpiomelli1999,anantharamu2020}.  In the smooth-wall two-component model of \citet{massey_eddy_2025} the $\ln\delta^+$ growth is carried by the nonlinear source over the inertial range \citep{massey2026nl}, while the linear source sets a near-wall offset \citep{massey2026inner}.

Over a rough wall the picture must change where the roughness changes the flow.  Once fully rough ($\ks^+\gg 1$, $\ks/\delta\ll 1$, with $\ks$ the equivalent sandgrain roughness) the viscous length is buried within the roughness sublayer and outer-layer similarity holds above it \citep{townsend1976,raupach1991,jimenez2004,schultz2007,squire2016}: the mean shear keeps its log-layer form $\mathrm{d}U^+/\mathrm{d}y_d^+ = 1/(\kappa y_d^+)$, with $\kappa$ the von K\'arm\'an constant and $y_d$ the wall-normal coordinate displaced by the zero-plane height (\S\ref{sec:framework}), shifted only by the roughness function.  Rough-wall pressure spectra have been measured and modelled extensively. \citet{blake1970} introduced roughness inner scaling for the high-frequency range; \citet{meyers2015} identified a three-region structure with a $\ks$-scaled intermediate band; and empirical Goody-type models \citep{goody2004,catlett2022,joseph2022,fritsch2023} extend the smooth-wall spectrum to rough walls, differing chiefly in which roughness parameter they carry.  A parallel theoretical line treats the rough-wall spectrum at low (acoustic and sub-convective) wavenumbers through enhanced source strengths and redistributed source near-fields \citep{howe1988,gleggdevenport2009}; it anticipates the source-density-times-Green-weighting partition used below but targets the scattering-dominated far field rather than the energetic convective peak, and we do not pursue it (cf.\ the sublayer-over-scattering assessment of \citealt{totten2026}).  What is missing is a simple account of what sets the growth of the variance once the wall is fully rough.

This work proposes that account and makes two contributions.  First, we carry the source-term interpretation of the wall-pressure spectrum to a fully rough wall, following its linear and nonlinear sources separately and asking how each rescales on its natural length: the linear source becomes roughness-local and scales on $\ks$, while the nonlinear source acts over a logarithmic wall-normal range now cut off below by the roughness sublayer rather than by the viscous scale.  Second, and as a direct corollary, the Reynolds-number growth of the integrated variance must count the geometric range $\eta_s\equiv\ln(\delta/\ks)$ in place of $\ln\delta^+$.  The reasoning is elementary: through the transitional band the viscous length $\nu/u_\tau$ and the roughness $\ks$ compete for the lower cutoff of the active source range, but once fully rough the viscous length is buried in the sublayer, $\ks^+$ loses independent content, and $\delta/\ks$ inherits the counting role of $\delta^+$.  This is the same outer-layer similarity that collapses the mean velocity profile, applied now to a range-counting quantity ---  a consequence of standard fully rough phenomenology, not a rival scaling.  The same similarity then fixes the rate: any canopy influence is set by $\ks$, so it decays in $y_d/\ks$ and integrates over the logarithmic measure to a constant, leaving the growth coefficient at its smooth-wall value and confining roughness to an additive offset.  Two premises are inherited and stated plainly at the outset: attributing the smooth-wall slope $A\approx 2.3$ to the nonlinear source over the log layer is imported from the smooth-wall analysis of \citet{massey2026nl}, and the split of the spectrum into a linear and a nonlinear part is an interpretation, not something the wall pressure measures directly.

The wall-pressure spectrum can be read as the wall imprint of these two mechanisms.  Over a rough wall the mean shear is largest where the canopy lifts the velocity onto its log profile, so $\Lin$ is concentrated in the roughness sublayer, scales on the roughness height, and produces a spectral feature at roughness-set frequencies.  The growth-relevant amplitude of $\NL$ lives over the inertial region above the sublayer, and the $y^{-1}$ decay there supplies the energetic outer peak at frequencies set by $\delta$ \citep{massey_eddy_2025,massey2026nl}.  The viscous (Stokes) pressure also contributes through the treatment of the boundary condition, but is subdominant, as discussed in \S\ref{sec:framework}. Figure~\ref{fig:schematic} sketches this two-population structure and how the Reynolds-number growth attaches to the nonlinear population alone.

\section{Source-term framework}\label{sec:framework}
In smooth-wall canonical flows, source-term analyses associate the near-wall $O(1)$ offset of the wall-pressure variance primarily with the linear pressure source $\Lin$ and the inertial-layer contribution with the nonlinear source $\NL$ \citep{kim1989,changpiomelli1999,panton2017,anantharamu2020,massey2026nl,massey2026inner}; \citet{changpiomelli1999} computed exactly this Green-weighted linear/nonlinear partition in channel flow.  Here we posit how each source rescales in a fully rough turbulent boundary layer when roughness is represented by the equivalent sandgrain roughness $\ks$, and use that scaling to construct a closed-form two-lognormal premultiplied spectrum with predicted break frequencies.

Let the roughness geometry be fixed, let $\ks^+\gg 1$, and assume $\ks/\delta\ll 1$.  The displaced wall-normal coordinate is $y_d=y-d$, where $d$ is the displacement height, with $x_2\equiv y_d$ in the sources of \eqref{eq:poisson}.  Above the roughness sublayer the mean field reduces to $\bar U_i=U(y_d)\,\delta_{i1}$, so $\Lin = -2(dU/dy_d)\,\partial_x v'$; using incompressibility the nonlinear source admits the local form $\NL = -(\partial_j u'_k)(\partial_k u'_j) = \tfrac12\omega_i\omega_i - s_{ij}s_{ij}$, the difference between fluctuating enstrophy and strain magnitude squared, with $\omega_i$ the fluctuating vorticity and $s_{ij}$ the fluctuating strain rate.  The two forms of $\NL$ differ by the non-fluctuating term $\partial_i\partial_j\overline{u'_iu'_j}$, which contributes only to the mean pressure.

For a wall-parallel wavenumber vector $\kvec=(k_x,k_z)$ with magnitude $k=|\kvec|$, the wall pressure is the Green projection
\begin{equation}
    \widehat p'(\kvec,t)
    =\int_0^\delta G(k,y_d)
    \left[\widehat{\Lin}(\kvec,y_d,t)+\widehat{\NL}(\kvec,y_d,t)\right]
    \dd y_d,
    \qquad
    G(k,y_d)\sim -\frac{e^{-k y_d}}{k},
\end{equation}
in the single-wall or large-$k\delta$ limit.  Throughout, $\widehat p'=\pi'_{y=0}$ is the wall trace of the pressure fluctuation, and superscript $+$ denotes viscous scaling ($\overline{p^{\prime 2}}^+\equiv\overline{p^{\prime 2}}/u_\tau^4$ for the kinematic variable, and likewise the premultiplied spectrum ${f\phi_{pp}}^+$, with $\phi_{pp}(f)$ the single-point frequency spectrum of the wall pressure at frequency $f$, normalised so that $\overline{p^{\prime 2}}=\int_0^\infty\phi_{pp}\,\dd f$).  The wall signal is a Green-weighted integral of wall-normal source densities; its variance involves $|G|^2 \sim e^{-2k y_d}/k^2$, so sources at height $y_d$ contribute to the wall with a factor that decays exponentially in $k y_d$.  What is used below is this wall-normal attenuation factor $e^{-k y_d}$, the dependence on $k y_d$ that selects which sources reach the wall; we do not rely on its accuracy as a $k\delta\to\infty$ approximation at the energetic peak, where $k\delta\sim O(1)$.  This elliptic attenuation is the central mechanism by which the wall-normal location of each source, and any wallward redistribution imposed by the canopy, enters the wall-projected variance.

Two simplifications warrant brief comment.  First, the mean field $\bar U_i$ remains three-dimensional within the canopy; a triple decomposition (time mean, steady spatial disturbance $\tilde U_i$, turbulence) would add a dispersive linear source, non-zero only within and just above the roughness sublayer \citep{raupach1982,mehdi2013,aghaeijouybari2022}.  It sits where the roughness-local linear contribution already sits and is absorbed into the same fitted constants; it does not reach the nonlinear population over its log-layer range $y_d\gg\ks$, where $\tilde U_i\to 0$.

Second, the half-space kernel $G\sim -e^{-ky_d}/k$ assumes a smooth wall at $y_d=0$, whereas here that plane lies within the canopy.  \citet{kim1989} found the analogous Stokes pressure negligible for the wall r.m.s.\ in smooth-wall channel flow; transferring the boundary condition through a Taylor expansion, the canopy correction is $O(k\ks)$ --- negligible for the energetic population at $k\sim 1/\delta$ ($k\ks\sim\ks/\delta\ll 1$), but $O(1)$ at roughness-scale wavenumbers, where it too is absorbed into the fitted roughness-local constants.

\input{figs/model_schematic}

\section{Linear source over a fully rough wall}\label{sec:linear}
Above the roughness sublayer the displaced mean velocity keeps the smooth-wall logarithmic form, shifted down only by the roughness function $\Delta U^+$.  This persistence of the log law above the roughness is Townsend's outer-layer similarity hypothesis \citep{townsend1976}, well supported for the mean flow across and below the present roughness range \citep{schultz2007,mehdi2013},
\begin{equation}
    U^+(y_d^+) \approx \kappa^{-1}\ln y_d^+ + B - \Delta U^+,
    \qquad
    \Delta U^+\approx \kappa^{-1}\ln \ks^+ + B_s,
\end{equation}
the roughness function $\Delta U^+$ being the downward shift of the log law with roughness, in the form first formulated by \citet{hama1954}.  Equivalently $U^+(y_d) \approx \kappa^{-1}\ln(y_d/\ks)+B_r$, with $B$ the smooth-wall log-law intercept, $B_s$ the fully rough intercept constant, and $B_r\equiv B-B_s$. In either representation,
\begin{equation}
    \frac{\partial U^+}{\partial y_d^+}\approx \frac{1}{\kappa y_d^+},
\end{equation}
so that the log-region mean shear has the same functional form as on the smooth wall, shifted only by the additive roughness function $\Delta U^+$.  Because $\Delta U^+$ enters $U$ but not $dU/dy_d$, it does not by itself introduce any new $\delta^+$ dependence into the log-layer part of $\Lin=-2(dU/dy_d)\,\partial_x v'$.  A leading change in the $\delta^+$ scaling of the linear contribution would therefore have to come from altered wall-parallel spectral weights, altered convection, or altered Green-projection into the one-dimensional frequency spectrum, not from the displaced logarithmic mean shear alone.  In the absence of those spectral changes, the linear-source contribution to the inner-scaled wall-pressure variance over the rough wall (subscript $r$) reduces to a constant,
\begin{equation}
    \langle p_{\Lin}^{\prime 2}\rangle_r^+
    \approx B_{\Lin}^{r}+O(1)
\end{equation}
as $\delta^+\to\infty$ at fixed fully rough state.  The constant $B_{\Lin}^{r}$ is roughness dependent but is not expected to grow logarithmically with $\delta^+$ or with $\delta/\ks$.

The important linear modification is local.  Within the roughness sublayer the mean field is three-dimensional (crest-scale gradients, sheltering, separation/reattachment) and the linear source couples those to fluctuations, giving a roughness-coherent contribution that replaces the smooth-wall buffer-layer feature.  In the reduced model this is parameterised by $\ks$: a roughness-scaled period $T^s\equiv TU_e/\ks$, with $T=1/f$ the period and $U_e$ the boundary-layer edge velocity, locates the peak and its amplitude $B_{\Lin}^{r}$ is a single fitted constant.  This $\ks$-scaled axis follows from the work of \citet{blake1970} who first proposed roughness inner scaling $\omega\ks/u_\tau$ for the high-frequency rough-wall spectrum.  Geometry sensitivity (element shape, solidity, sheltering, the dispersive source of \S\ref{sec:framework}, the canopy-to-wall projection) is absorbed into $B_{\Lin}^{r}$, $T_{b\Lin}^{s,r}$, $\sigma_{\Lin}^r$ for the Virginia Tech (VT) geometry of \S\ref{sec:vt}; the assumption is that these constants, whatever their values, do not grow with $\delta/\ks$ at fixed roughness state.

\section{Nonlinear source over a fully rough wall}\label{sec:nonlinear}

The nonlinear source $\NL$ is the dominant inertial-layer pressure mechanism in the smooth-wall source-term interpretation \citep{massey2026nl,massey2026inner}.  Over a fully rough wall the active wall-normal range of $\NL$ is cut off below by the roughness sublayer rather than by the viscous scale, so the natural log-range parameter is
\begin{equation}
    \eta_s \equiv \ln\!\left(\frac{\delta}{\ks}\right),
\end{equation}
which sets the outer extent of the fully rough logarithmic source region in place of $\ln\delta^+$ on the smooth wall.  This is the pressure-variance counterpart of the rough-wall outer-layer similarity measured for the velocity statistics \citep{squire2016}: above the sublayer the mean shear and energetic turbulence take their smooth-wall log form, so the nonlinear source density the variance integrates does too, now over $[\ks,\delta]$ rather than $[\nu/u_\tau,\delta]$.  At leading order in the fully rough regime, $\ks^+$ does not enter $\NL$ as an independent dynamical parameter: the viscous length is hidden inside the roughness sublayer and the active range depends on $\delta$ and $\ks$ only through their ratio.

A second effect is entirely about the wall-normal position of the sources, all of which sit at $y_d>0$ above the displacement plane.  The elliptic kernel attenuates each source by $e^{-ky_d}$, so a source deeper in the layer (larger $y_d$) reaches the wall more weakly.  ``Wallward'' therefore means toward smaller $y_d$: if the canopy holds the active nonlinear source closer to the wall than its smooth-wall counterpart sits, by an effective amount $\Delta_{\NL}(y_d)>0$, that source is less attenuated and its wall imprint rises.  Collect that departure in a single local gain,
\begin{equation}
    \chi_G(y_d)=\exp\!\left[2 k_*(y_d)\,\Delta_{\NL}(y_d)\right],
    \qquad
    \chi_G\to 1 \quad\text{as } y_d/\ks\to\infty,
    \label{eq:green_gain_local}
\end{equation}
so that $\chi_G\equiv 1$ for the smooth wall and $\chi_G>1$ for a wallward shift, with $k_*(y_d)$ the wall-parallel wavenumber most relevant to the source at $y_d$.  Nothing in the construction fixes the sign of the shift: outward redistribution ($\Delta_{\NL}<0$, $\chi_G<1$) is equally admissible.  The limit in \eqref{eq:green_gain_local} is not an extra assumption.  The canopy has one length, $\ks$, and above the sublayer outer-layer similarity leaves no other, so $\chi_G-1$ is a function of $y_d/\ks$ alone and must vanish as that ratio grows.

Splitting the wall-projected nonlinear variance at the roughness height and writing $\chi_G=1+(\chi_G-1)$ in the outer part gives
\begin{equation}
    \langle p_{\NL}^{\prime 2}\rangle_r^+ \sim
    \underbrace{\int_{y_{d,0}}^{\ks} a_{\NL}^{\mathrm{src},r}(y_d)\,
    \chi_G(y_d)\,\dd\ln y_d}_{\displaystyle \equiv\, B_{\NL,i}^r}
    \;+\; A_{\NL,\eta}^{\mathrm{base}}
    \left[\,\eta_s+\int_{1}^{\delta/\ks}\bigl(\chi_G-1\bigr)\,\dd\ln \frac{y_d}{\ks}\,\right],
    \label{eq:nonlinear_integral}
\end{equation}
with $y_{d,0}$ a finite inner floor at the element scale that, with the rest of the sublayer contribution, is absorbed into $B_{\NL,i}^r$.  The first term is the inner-scaled invariant part of $\NL$, generated within the roughness sublayer (crest-scale turbulence, separation and reattachment, the canopy-coherent motions of rough-channel DNS; \citealt{bhaganagar2007}).  In the second term $a_{\NL}^{\mathrm{src},r}$ has been set to $A_{\NL,\eta}^{\mathrm{base}}$, its universal log-layer value: outer-layer similarity supplies the source density above the sublayer (the attached-eddy/uniform-momentum-zone baseline, supported on rough walls by the outer-layer UMZ similarity of \citealt{zhang2026}), and roughness does not renormalise it.

Because $\chi_G-1$ decays in ${y_d/\ks}$, the remaining integral converges,
\begin{equation}
    \int_{1}^{\delta/\ks}\bigl(\chi_G-1\bigr)\,\dd\ln \frac{y_d}{\ks}
    \;\longrightarrow\;
    \mathcal I_\chi\equiv\int_{1}^{\infty}\bigl(\chi_G-1\bigr)\,\dd\ln \frac{y_d}{\ks}
    \qquad (\delta/\ks\to\infty),
    \label{eq:Ichi}
\end{equation}
to a number set by the roughness geometry and independent of $\delta/\ks$.  Hence
\begin{equation}
    \langle p_{\NL}^{\prime 2}\rangle_r^+
    \approx A_{\NL,\eta}^{\mathrm{base}}\,\eta_s + B_{\NL,G}^r,
    \qquad
    B_{\NL,G}^r \equiv B_{\NL,i}^r + A_{\NL,\eta}^{\mathrm{base}}\,\mathcal I_\chi,
    \label{eq:nonlinear_growth}
\end{equation}
the smooth-wall coefficient on the rough-wall range, with every roughness-set effect in a single non-growing offset.

This subsumes the loose ends of \S\ref{sec:framework} under one principle.  The dispersive linear source, the $O(k\ks)$ canopy correction to the half-space kernel, the sublayer-generated nonlinear turbulence and any Green-weighted redistribution of the log-layer source are all set by the roughness and all die away with $y_d/\ks$.  Each therefore contributes a convergent, $\eta_s$-independent amount to $B_{\NL,G}^r$, and none of them can alter the coefficient of $\eta_s$.  The wall spectrum constrains only their sum; separating them requires wall-normal source statistics (\S\ref{sec:discussion}).

Framed as a falsifiable test, a multiplicative gain on the growth rate would require a canopy influence that persists undiminished to the top of the layer --- $\Delta_{\NL}\propto y_d$ all the way to $y_d\sim\delta$, so that roughness of height $\ks\sim\delta/30$ displaces the mid-layer source by $O(\delta)$.  That is what outer-layer similarity forbids, and it is the same similarity that produced $\eta_s$.  The growth rate is therefore not a free parameter of the model but a prediction, and a measured rate departing from $A_{\NL,\eta}^{\mathrm{base}}$ is evidence against outer-layer similarity for the pressure source---or against attributing the whole smooth-wall rate to $\NL$ (\S\ref{sec:model})---rather than a canopy gain to be absorbed.

\section{Two-lognormal rough-wall model}\label{sec:model}
    The smooth-wall two-component model writes the premultiplied spectrum (at sufficient $\delta^+$) as an inner population plus an outer population \citep{massey_eddy_2025},
    \begin{equation}
        {{f\phi_{pp}}}^{+}=g_1(T^+)+g_2(T^o;\delta^+),
        \qquad
        T=1/f,
        \qquad
        T^+\equiv\frac{Tu_\tau^2}{\nu},
        \qquad
        T^o=\frac{T\ue}{\delta}.
    \end{equation}
    For fully rough walls, splitting the spectrum by its two sources gives
    \begin{equation}
        {f\phi_{pp,r}}^{+}
        = g_{\Lin}^r(T^s;\mathcal R_s)+g_{\NL}^r(T^o;\eta_s,\mathcal R_s),
        \qquad
        T^s\equiv\frac{T\ue}{\ks},
        \qquad
        \eta_s\equiv\ln\!\left(\frac{\delta}{\ks}\right),
    \end{equation}
    where $\mathcal R_s$ denotes the fixed fully rough hydrodynamic state (absorbing all geometry-set constants).\footnote{Both populations use the measured edge velocity $\ue$, which keeps both breaks on one directly measurable axis; a canopy velocity $U_d<\ue$ for the roughness-local feature would rescale $T_{b\Lin}^{s,r}$ by the constant $\ue/U_d$ and is absorbed into the fitted constant (cf.\ \eqref{eq:linear_shed}).}  Only two lognormal populations are retained, and $\eta_s$ is the sole carrier of Reynolds-number growth at fixed roughness state. The closest spectral-model antecedent is the three-region rough-wall structure of \citet{meyers2015}, built on the same VT-lineage measurements; the present model differs in deriving its two populations from the two Poisson source terms, carrying all Reynolds-number growth in the outer area through $\eta_s$, and retaining no viscous-scaled range once fully rough.  Define the lognormal-in-$T$ density
    \begin{equation}
        \mathcal N(T;T_b,\sigma,A)
        =\frac{A}{\sqrt{2\pi}\sigma}
        \exp\!\left[-\frac{1}{2}\!\left(\frac{\ln(T/T_b)}{\sigma}\right)^{\!2}\right],
    \end{equation}
    so that $\int_0^\infty \mathcal N\,\dd\ln f = A$ over an infinite logarithmic frequency domain.  Then
    \begin{align}
        g_{\Lin}^r(T^s)
        &= \mathcal N\!\left(T^s;T_{b\Lin}^{s,r},
        \sigma_{\Lin}^r,B_{\Lin}^r\right),
        \\
        g_{\NL}^r(T^o)
        &= \mathcal N\!\left(T^o;T_{b\NL}^{o,r},
        \sigma_{\NL}^r,A_{\NL,\mathrm{obs}}^r\right),
    \end{align}
    with the nonlinear area set by \eqref{eq:nonlinear_growth},
    \begin{equation}
        A_{\NL,\mathrm{obs}}^r
        = A_{\NL,\eta}^{\mathrm{base}}\,\eta_s + B_{\NL,G}^r,
        \qquad A_{\NL,\eta}^{\mathrm{base}}\approx 2.3 ,
        \label{eq:gq_area}
    \end{equation}
    with the intercept collecting every non-growing nonlinear contribution of \S\ref{sec:nonlinear} --- the inner-scaled sublayer offset and the convergent canopy Green term.  Here $A_{\NL,\eta}^{\mathrm{base}}\equiv a_{\NL}^{\mathrm{src},r}$ is the bare log-range coefficient set by the universal log-layer nonlinear-source density.  We take $A_{\NL,\eta}^{\mathrm{base}}=A_{\NL}^{\mathrm{smooth}}\approx 2.3$, the natural-log slope of $\overline{p^{\prime 2}}^+$ against $\ln\delta^+$ in the smooth-wall data of \citet{schlatterorlu2010,leemoser2015} (fitted slopes $2.42$ and $2.24$). 

    Attributing the entire smooth slope to $\NL$ takes the linear source to carry no $\ln\delta^+$ growth --- the partition \citet{massey2026inner} measure on the smooth wall, against the classical attribution of the overlap-range spectrum, and hence part of the growth, to the mean-shear (linear) source \citep{bradshaw1967,pantonlinebarger1974}.  The rate prediction is less sensitive to this partition than it may appear: the log-layer mean shear keeps its smooth-wall form above the sublayer (\S\ref{sec:linear}), so any log-layer linear growth would count on $\eta_s$ alongside $\NL$ and leave the total coefficient at $\approx 2.3$; the partition fixes only which source carries the growth, an attribution a wall-pressure sweep cannot resolve (\S\ref{sec:nonlinear}).  No $\ks^+$-dependent term appears in \eqref{eq:gq_area}.

\section{Break frequencies}\label{sec:breaks}
The break frequency of each population follows from where its source sits and how fast that source is carried past the wall.  For the roughness-local linear population the break sits on the roughness period axis $T^s$ and, at fixed roughness state, should not drift with $\delta^+$:
\begin{equation}
    f_{b\Lin}^{r}\sim \frac{U_{c,\Lin}^{r}}{C_{b\Lin}^r\ks},
    \qquad
    T_{b\Lin}^{s,r}\equiv
    \frac{\ue}{f_{b\Lin}^{r}\ks}
    \sim C_{b\Lin}^r\frac{\ue}{U_{c,\Lin}^{r}},
    \label{eq:linear_break}
\end{equation}
with $U_{c,\Lin}^{r}$ the convection velocity of the roughness-local motions and $C_{b\Lin}^r=O(1)$.  What sets the \emph{numerical} value of the break is the physical mechanism behind the feature, and two natural candidates give predictions that differ by about a factor of five --- a gap wide enough that the fit of \S\ref{sec:vt} can tell them apart.

The first candidate is unsteady vortex shedding from the roughness elements.  The VT bed is staggered circular cylinders of diameter $d_c=3.14$~mm \citep{totten2026}, and a cylinder in cross-flow sheds a K\'arm\'an vortex street at the Strouhal frequency $f_s=St_{d_c}\,U/d_c$, with $St_{d_c}\approx 0.2$ over a broad Reynolds-number range.  The approach speed the elements feel is not the edge velocity but the reduced canopy velocity $U_d\approx(0.4$--$0.5)\ue$ they sit in low in the layer, so $f_s=St_{d_c}\,U_d/d_c$ and, on the roughness period axis,
\begin{equation}
    T_{b\Lin}^{s,r}=\frac{\ue}{f_s\ks}
    =\frac{1}{St_{d_c}}\,\frac{\ue}{U_d}\,\frac{d_c}{\ks},
    \label{eq:linear_shed}
\end{equation}
fixed by the geometric ratio $d_c/\ks$ and the convection deficit, not by $\delta^+$.  With $St_{d_c}\approx 0.2$, $U_d/\ue\approx 0.4$--$0.5$ and $d_c/\ks\approx 0.95$, \eqref{eq:linear_shed} predicts $T_{b\Lin}^{s,r}\approx 10$--$12$.  Shedding is a slow dynamical instability, so it would place the feature at long period, well to the low-frequency (large-$T^s$) side of the roughness axis.

The second candidate is simpler: an element-scale structure convected past the wall at the roughness convection velocity, with no instability frequency of its own.  Using the measured $U_c\approx 0.49\,\ue$ \citep{totten2026} and a streamwise wavelength of order the element size, $\lambda_x\sim\ks$, this feature falls at $T_{b\Lin}^{s,r}=\ue\lambda_x/(U_c\ks)\sim\ue/U_c\approx 2$ --- of order unity on the roughness axis, a factor of about five below the shedding value.  The two mechanisms are thus cleanly separated on $T^s$: shedding near $T^s\approx 10$--$12$, convected element-scale structure near $T^s\sim 2$.  Which of the two the data select is left to the fit of \S\ref{sec:vt}.

For the nonlinear population, let the wall-effective active source height be $y_{\NL,\mathrm{eff}}\gtrsim\ks$.  Then
\begin{equation}
    T_{b\NL}^{o,r}\sim
    C_{b\NL}^r\,
    \frac{\ue}{U_{c,\NL}^{r}(y_{\NL,\mathrm{eff}})}
    \frac{y_{\NL,\mathrm{eff}}}{\delta}.
    \label{eq:nonlinear_break}
\end{equation}
Here $U_{c,\NL}^{r}(y_{\NL,\mathrm{eff}})$ is the convection velocity of the energetic nonlinear motions at that height and $C_{b\NL}^r$ an $O(1)$ prefactor.  At fixed roughness state an increase in $\delta/\ks$ should grow the area of $g_{\NL}^r$ through \eqref{eq:gq_area} without moving the break; a break shift in $T^o$ would instead signal a change in $y_{\NL,\mathrm{eff}}/\delta$, $U_{c,\NL}^r/\ue$, or the Green-weighted source distribution.

The prefactor is not pinned by the construction, and the inferred source height depends on it inversely ($y_{\NL,\mathrm{eff}}/\delta = T_{b\NL}^{o,r}(U_{c,\NL}^r/\ue)/C_{b\NL}^r$); the elliptic kernel constrains the admissible combinations.  At the energetic peak $T^o\approx 1$ with $U_c\approx 0.5\,\ue$ the convected wavelength is $\lambda_x\approx 0.5\,\delta$, and a source at $y_d=0.5\,\delta$ ($C_{b\NL}^r=1$) has $ky_d\approx 2\pi$ and negligible kernel weight $e^{-4\pi}$ at the wall; an $O(1)$ kernel ($ky_d\lesssim 1$) instead puts the source in the log layer, $y_{\NL,\mathrm{eff}}/\delta\lesssim 0.1$, with $C_{b\NL}^r=O(2\pi)$ absorbing the aspect ratio of wall-attached structures.  The wall spectrum alone does not distinguish the two readings; \S\,\ref{sec:vt} reports both.

\section{Model results at the VT cases}\label{sec:vt}
The model of \S\ref{sec:model} is evaluated at fourteen rough-wall cases from the Virginia Tech Stability Wind Tunnel rough-wall campaign \citep{vishwanathan2023,fritsch2023}: a single cylinder-roughness plate (staggered cylinders, $d_c=3.14$~mm, $k_g=2$~mm, pitch $6.93$~mm; \citealt{totten2026}) at nominally zero pressure gradient (the tunnel's NACA~0012 at zero angle of attack), sampled at seven streamwise stations $x=1.25$--$4.91$~m in each of two unit-Reynolds-number families (nominal $\ue=35$ and $65$~m/s; measured $\ue\approx 32$--$36$ and $57$--$65$ across stations): a spatially developing fully rough boundary layer measured along its growth.  The measured flow state ($Re_\tau\equiv\delta^+$, $\ue$, $\delta$, $u_\tau$, $C_f$) is supplied for each case; the roughness is represented by the equivalent sandgrain height $\ks=3.3$~mm, the value implied by the dataset's own roughness-scaled spectral axes and some $3\%$ above the $\ks\approx 1.6k_g$ quoted by \citet{totten2026}, whose sub-convective measurements on the same roughness lineage also supply the convection velocity used below.  Every case is fully rough, $\ks^+\in[361,702]$, with $\delta/\ks\in[18,42]$ --- low enough that Townsend similarity is an applied premise \citep{jimenez2004}, though one the boundary-layer evidence supports down to this range \citep{schultz2007,squire2016}.  The model treats all fourteen as canonical fully rough states parameterised by $(\delta^+,\ks)$ alone.

\subsection{Calibration}
The six parameters are estimated in two separate stages.  It is worth being explicit that all fourteen cases are used to calibrate: with a single roughness geometry there is no independent set to validate against, so the model is calibrated rather than cross-validated, and its honesty rests on an error bar that reflects how few independent measurements the data really contain (below), not on a train/test split the data cannot support.

Stage~A (shape) fits the four shape parameters $(T_{b\Lin}^{s,r},\sigma_{\Lin}^r,T_{b\NL}^{o,r},\sigma_{\NL}^r)$ and the linear area $B_{\Lin}^r$ by bounded least squares to the area-normalised spectra ${f\phi_{pp}}^+/\overline{p^{\prime 2}}^+$ ($2221$ spectral points across the fourteen cases), where $\overline{p^{\prime 2}}^+=\int {f\phi_{pp}}^+\,\dd\ln f$ is each case's measured integrated variance.  Dividing each spectrum by its own variance removes the $\pm 40\%$ amplitude scatter that the campaign's $\pm 20\%$ $C_f$ uncertainty imposes, leaving only the shape; the non-growing linear population enters as a constant absolute area.  Stage~B (growth) then tests the predicted coefficient against the fourteen measured variances and calibrates the single remaining offset,
\begin{equation}
    \begin{aligned}
        \overline{p^{\prime 2}}^+ &= V_0 + A_{\NL,\eta}^r\,\eta_s
        &&\quad\text{(test, $A_{\NL,\eta}^r$ free)},\\
        \overline{p^{\prime 2}}^+ &= V_0 + A_{\NL,\eta}^{\mathrm{base}}\,\eta_s
        &&\quad\text{(calibration)},\\
        B_{\NL,G}^r &= V_0 - B_{\Lin}^r,
    \end{aligned}
    \label{eq:growth_regression}
\end{equation}
with $V_0$ the intercept and each case's $\eta_s=\ln(\delta/\ks)$ from its measured $\delta$; the slope reads as the nonlinear growth rate because the linear population does not grow.

Keeping the two stages apart matters because they rest on very different amounts of information.  The shape is pinned by $2221$ spectral samples freed of amplitude noise; the growth test and the offset rest on a single fourteen-point regression whose $\eta_s$ range is generated by streamwise development, so it inherits whatever non-equilibrium history and residual tunnel gradient the developing layer carries \citep{vishwanathan2023} and is an estimate, not a controlled measurement.  Uncertainties for both stages come from a case-level bootstrap: the fourteen cases are resampled with replacement, both stages refitted on each of $1000$ replicates, and the central $68\%$ spread reported (table~\ref{tab:params}).  Resampling whole cases --- not individual spectral points, which are strongly correlated within a case and would manufacture false precision --- keeps the effective sample size at the fourteen cases (arguably the two speed families) that vary independently.  This is why the shape intervals come out at a few percent while the free-slope test interval is wide.

\subsection{Outcome}
\begin{figure}
    \centering
    \includegraphics[width=\textwidth]{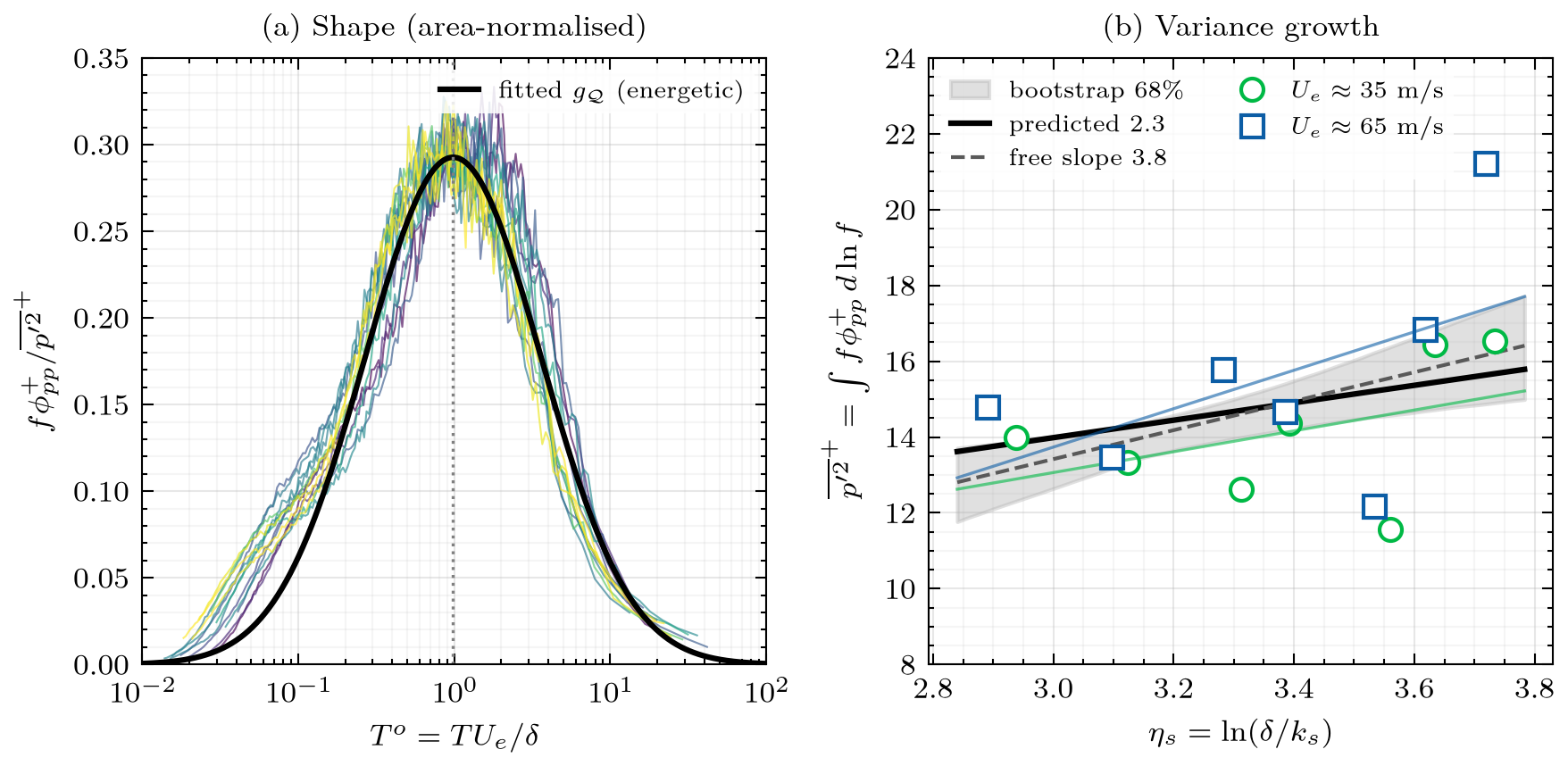}
    \caption{The decoupled calibration.  (a)~The fourteen area-normalised premultiplied spectra ${f\phi_{pp}}^+/\overline{p^{\prime 2}}^+$ on the outer axis $T^o$, coloured by $Re_\tau$, with the fitted energetic lognormal $g_{\NL}^r$ overlaid (heavy black, at the mean energetic area fraction $\approx 0.95$); the energetic peak collapses at $T^o\sim 1$ and the roughness-local population $g_{\Lin}^r$ appears as the small shoulder at low $T^o$.  (b)~Measured integrated variance $\overline{p^{\prime 2}}^+$ versus $\eta_s=\ln(\delta/\ks)$ for the two speed families, with the predicted slope $A_{\NL,\eta}^{\mathrm{base}}=2.3$ at the fitted offset \eqref{eq:growth_regression} (heavy black), the free-slope least-squares fit (dashed) and its bootstrap $68\%$ band (grey), and the family fits (thin).}
    \label{fig:shape_growth}
\end{figure}

The shape is tightly constrained; the free-slope test is not.  Stage~A collapses the fourteen area-normalised spectra onto a common shape at an r.m.s.\ residual of $0.02$, with bootstrap intervals a few percent of the central values (figure~\ref{fig:shape_growth}a, table~\ref{tab:params}).  Stage~B returns a free slope $A_{\NL,\eta}^r\approx 3.8$, but the variances scatter about the regression with $R^2\approx 0.19$: the bootstrap $68\%$ interval is $[1.5,\,6.1]$ and the $95\%$ interval $[-1.0,\,9.5]$ does not exclude zero slope (figure~\ref{fig:shape_growth}b).  The predicted $A_{\NL,\eta}^{\mathrm{base}}=2.3$ sits comfortably inside it, and imposing it costs almost nothing: the residual sum of squares rises by $3.6\%$ while one parameter is removed.  The VT data are therefore consistent with the predicted rate but cannot yet discriminate it from a rate half or twice as large.  Nor can they prefer the $\eta_s$ basis over $\ln\delta^+$: regressed on either, the fourteen variances scatter about the trend with the same $R^2\approx 0.19$, so the campaign does not favour one basis on fit quality.  The two are not the same regressor --- because $\ks^+$ varies across the campaign they are only moderately correlated ($\mathrm{corr}\approx 0.66$), spanning $\Delta\eta_s\approx 0.84$ against $\Delta\ln\delta^+\approx 1.20$ --- and what separates them is exactly the family offset $\Delta\ln\ks^+\approx 0.53$ between the two speed families at matched $\eta_s$; the choice of $\eta_s$ therefore rests on the outer-similarity argument of \S\ref{sec:nonlinear}, not on a data preference.  Separating them requires varying $\ks^+$ at matched $\delta/\ks$, and the two speed families are exactly that comparison, taken up below. The point estimate there lies nearer the $\ln\delta^+$ prediction than zero, though neither is excluded.

The rate is, moreover, hostage to single cases at the ends of so short a range (figure~\ref{fig:shape_growth}b).  The low pair near $\eta_s\approx 3.5$ --- the $35$ and $65$~m/s stations at $x=3.69$~m --- sits below the trend; removing the $35$~m/s member alone steepens the fitted slope from $3.8$ to $4.6$ and lifts $R^2$ from $0.19$ to $0.31$.  Removing instead the single highest point at the other end ($65$~m/s, $x=4.91$~m) flattens the slope to $1.7$.  Both bracket the predicted $2.3$, and no one deletion makes the rate robust; over fourteen points spanning only $\Delta\eta_s\approx 0.84$ none should.

A common-mode $C_f$ bias is degenerate with the prediction, since scaling all variances by $c$ tilts the test slope to $2.3c$.  Closing the gap from $2.3$ to $3.8$ requires $c=1.65$, that is $u_\tau$ overestimated by $13\%$ and $C_f$ by $28\%$ --- beyond the campaign's quoted $\pm 20\%$, so calibration bias at the stated level accounts for much but not all of the difference, the remainder lying well inside $[1.5,\,6.1]$.

What the free-slope test leaves loose the calibration pins: at the predicted slope the offset is well determined, $B_{\NL,G}^r\approx 6.4\pm 0.6$ ($68\%$).  Dividing it by the range coefficient converts it into the currency of the log range, $B_{\NL,G}^r/A_{\NL,\eta}^{\mathrm{base}} = \mathcal I_\chi + B_{\NL,i}^r/A_{\NL,\eta}^{\mathrm{base}}\approx 2.8$: the sublayer and canopy together add as much inner-scaled variance as a further factor $e^{2.8}\approx 16$ in $\delta/\ks$ would.  That is comparable to the whole range these layers possess ($\eta_s\approx 2.9$--$3.7$), so the non-growing offset supplies over $40\%$ of their variance --- consistent in sign with the enhanced near-wall pressure activity of rough-wall DNS \citep{bhaganagar2007}.  The outer break $T_{b\NL}^{o,r}\approx 0.98$ is likewise tight: with the measured $U_c\approx 0.49\,\ue$ \citep{totten2026} it fixes the convected wavelength $\lambda_x\approx 0.5\,\delta$, consistent with the smooth-wall outer population of \citet{massey_eddy_2025}, and via \eqref{eq:nonlinear_break} places the effective source height at $y_{\NL,\mathrm{eff}}/\delta\approx 0.5$ if $C_{b\NL}^r=1$ or $\lesssim 0.1$ under the kernel-consistent $C_{b\NL}^r=O(2\pi)$.

The fitted \emph{linear} break settles the mechanism left open in \S\ref{sec:breaks}.  It falls at $T_{b\Lin}^{s,r}\approx 1.87$ (table~\ref{tab:params}), on the convected element-scale prediction $\ue/U_c\approx 2$ and a factor of five to six below the shedding value $10$--$12$ of \eqref{eq:linear_shed}: the roughness-local feature is element-scale structure convected past the wall, not vortex shedding from the cylinders.  The axis relation $T^s=(\delta/\ks)\,T^o$ also reassigns a feature seen near $St\equiv f\ks/\ue\approx 0.02$ in the VT spectra, read tentatively in earlier work as a canopy `breathing' mode: it maps to $T^o\approx 1.2$--$2.8$ across these $\delta/\ks$, i.e.\ onto the energetic nonlinear population at $T^o\sim 1$ rather than a roughness-set mode, so the genuinely roughness-local content is the $g_{\Lin}^r$ feature at $T^s\sim 2$.

\begin{table}[h]
    \centering
    \small
    \setlength{\tabcolsep}{4pt}
    \begin{tabular}{llllr}
        \hline
        Parameter & Symbol & Value & 68\% interval & Source\\
        \hline
        \multicolumn{5}{l}{Stage A: spectral shape (area-normalised spectra)}\\
        Linear break (roughness scaling)  & $T_{b\Lin}^{s,r}$ & 1.87  & $[1.81,\,1.96]$ & fitted \\
        Linear width                      & $\sigma_{\Lin}^r$ & 0.50  & $[0.48,\,0.52]$ & fitted \\
        Linear area                       & $B_{\Lin}^r$      & 0.70  & $[0.67,\,0.78]$ & fitted \\
        Nonlinear break (outer scaling)   & $T_{b\NL}^{o,r}$  & 0.98  & $[0.94,\,1.04]$ & fitted \\
        Nonlinear width                   & $\sigma_{\NL}^r$  & 1.297 & $[1.280,\,1.304]$ & fitted \\
        \multicolumn{5}{l}{Stage B: variance growth (14-point regression on $\eta_s$)}\\
        Fully rough range coefficient     & $A_{\NL,\eta}^{\mathrm{base}}$ & 2.30 & --- & \emph{predicted} \\
        \quad free-slope fit (test)       & $A_{\NL,\eta}^r$  & 3.8   & $[1.5,\,6.1]$   & regression slope \\
        Nonlinear baseline                & $B_{\NL,G}^r$     & 6.4   & $[5.7,\,6.9]$   & $V_0-B_{\Lin}^r$ \\
        \hline
    \end{tabular}
    \caption{Two-lognormal rough-wall parameters from the two stages of \S\ref{sec:vt}.  The range coefficient is predicted, not fitted; the free-slope fit is shown as the test of it.  Intervals are central $68\%$ ranges from a case-level bootstrap ($1000$ replicates, whole cases) and quantify sampling scatter only: a common-mode $C_f$ calibration bias ($\pm 20\%$) would rescale the free-slope test and the offset by up to $\approx\pm 40\%$ without registering here.  Shape parameters and $B_{\NL,G}^r$ are specific to the VT geometry.}
    \label{tab:params}
\end{table}

\begin{figure}
    \centering
    \includegraphics[width=0.9\textwidth]{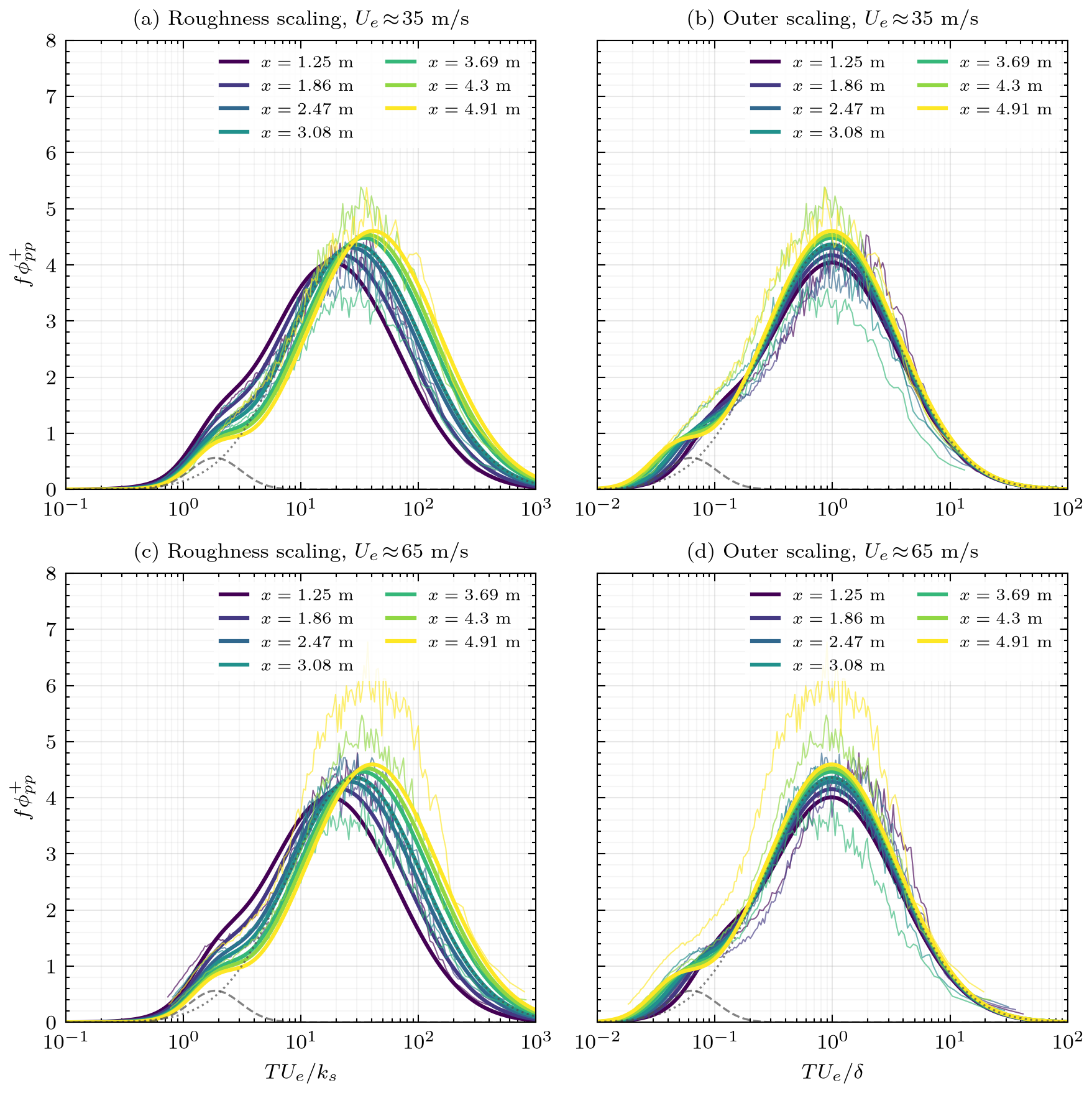}
    \caption{Calibrated two-lognormal model (heavy coloured curves) overlaid on the VT premultiplied spectra (thin coloured lines), in roughness-scaled ($TU_e/\ks$) and outer-scaled ($T^o=T\ue/\delta$) coordinates, at the parameter values of table~\ref{tab:params} and each case's measured flow state.  Dashed and dotted grey curves show $g_{\Lin}^r$ and $g_{\NL}^r$ individually for a representative case in each row.}
    \label{fig:vt_data_fit}
\end{figure}

Figure~\ref{fig:vt_data_fit} overlays the calibrated model on the VT spectra in roughness-scaled and outer-scaled coordinates, evaluated at the measured per-case flow states.  The energetic peak collapses on the outer axis at $T^o\sim 1$, the roughness-local feature sits near $T^s\sim 2$, and the predicted peak and variance rise across the sweep ($\max {f\phi_{pp}}^+\approx 4.0\to 4.5$ and $\overline{p^{\prime 2}}^+\approx 13.8\to 15.4$ from $Re_\tau\approx 7400$ to $24600$), inside the VT envelopes.  The r.m.s.\ residual of the assembled model is $0.50$ on the dimensional ${f\phi_{pp}}^+$ (peaks $\sim 4$--$7$); the $0.02$ of Stage~A is on the unit-area shape.

The two speed families overlap in $\eta_s$ at distinctly different $\ks^+$ (family means $\approx 393$ and $669$) and so directly probe whether $\ks^+$ drops out once fully rough.  The probe is inconclusive.  The families return different fitted slopes ($A_{\NL,\eta}^r\approx 2.75$ at $35$~m/s, $5.07$ at $65$~m/s), but a common-slope fit with a family offset gives $+1.5\pm 1.2$ at matched $\eta_s$ --- within $1.3$ standard errors of zero (the $\eta_s$ basis) and within a third of a standard error of the $A_{\NL,\eta}^{\mathrm{base}}\,\Delta\ln\ks^+\approx 1.2$ the $\ln\delta^+$ basis predicts ($\Delta\ln\ks^+\approx 0.53$ between families) --- and the split hangs on the same single case as above: the high-variance $65$~m/s point at $x=4.91$~m, whose removal drops the $65$~m/s family slope to $0.8$.  Genuine residual $\ks^+$ sensitivity, non-equilibrium development effects, a family-level $C_f$ bias ($\sim 5\%$ suffices) and experimental scatter are not separable from the VT data alone (\S\ref{sec:discussion}).

\section{Smooth-to-rough transition}\label{sec:transition}
The same two-source construction sketches how the spectrum reorganises as the wall is taken from hydraulically smooth, through the transitional band ($5\lesssim\ks^+\lesssim 70$), into the fully rough regime; only the two asymptotic limits are fixed by the framework, the band between them requiring an interpolating ansatz it does not derive.  In the smooth limit the inner population sits at the viscous peak $T^+\sim 20$ and the outer population grows as $\ln\delta^+$ at the rate $A_{\NL}^{\mathrm{smooth}}\approx 2.3$; in the fully rough limit the viscous inner peak is replaced by the roughness-local linear population $g_{\Lin}^r(T^s)$, the inner-scaled invariant nonlinear part contributes the offset $B_{\NL,i}^r$, and the outer population grows instead on $\eta_s$.  Through the band the viscous peak is hollowed out while a roughness-local feature fills the same viscous decade, and the outer peak, fixed in $T^o$, only changes amplitude; a peak that appears to move in viscous units across roughness states is therefore an artefact of the rescaling, not of a moving source.

\begin{figure}
    \centering
    \includegraphics[width=\textwidth]{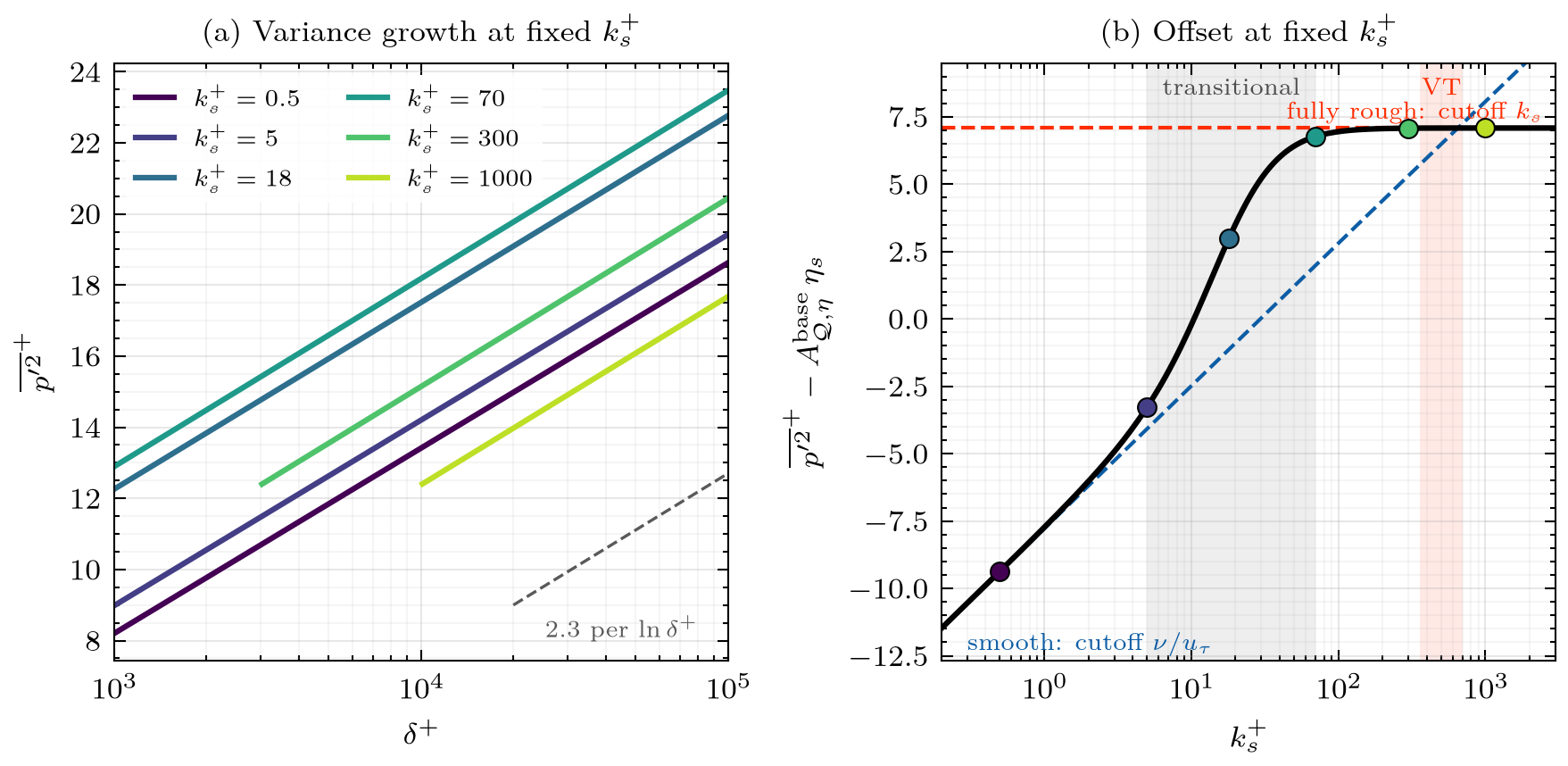}
    \caption{Effect of $\ks^+$ on the logarithmic growth of the wall-pressure variance, from a smooth-to-rough blend of the present model at fixed $C_f$.  (a)~Integrated variance $\overline{p^{\prime 2}}^+$ versus $\delta^+$ at six fixed $\ks^+$ from hydraulically smooth ($\ks^+=0.5$) to fully rough ($\ks^+=1000$), drawn only where $\delta/\ks\ge 10$; the curves are parallel (grey guide, $2.3$ per $\ln\delta^+$), roughness entering as a vertical offset alone.  (b)~That offset, $\overline{p^{\prime 2}}^+-A_{\NL,\eta}^{\mathrm{base}}\eta_s$, at fixed $\ks^+$: it rises as $A_{\NL,\eta}^{\mathrm{base}}\ln\ks^+$ while the lower cutoff of the active range is still $\nu/u_\tau$ (blue dashed), and saturates at the geometry-set $B_{\Lin}^r+B_{\NL,G}^r$ (red dashed) once $\ks$ takes over, i.e.\ once $\ks^+\gtrsim O(100)$, so the VT cases (red band) sit on the plateau.  The grey band marks where the cutoff migrates from $\nu/u_\tau$ to $\ks$.  Only the two limits are fixed by the framework; the width and shape of the crossover are the blend's.}
    \label{fig:growth_coeff}
\end{figure}

This yields a structural prediction sharper than any fitted number (figure~\ref{fig:growth_coeff}).  At fixed $\ks^+$ the inner-scaled variance grows linearly in $\ln\delta^+$ at the coefficient $2.3$ throughout, because $\eta_s=\ln\delta^+-\ln\ks^+$ and $\ks^+$ is fixed: roughness enters the growth not through the rate but through a vertical offset.  The transitional band is traversed not along these curves but across them, by changing $\ks^+$: the offset rises as $A_{\NL,\eta}^{\mathrm{base}}\ln\ks^+$ while the lower cutoff of the active range is still $\nu/u_\tau$, and saturates at a geometry-set constant once $\ks$ takes over.  It is that crossover --- not any change of rate at fixed $\ks^+$ --- whose width and shape are properties of the chosen blend rather than of the derivation.  A $\delta^+$ sweep at fixed $\ks^+$ in the fully rough regime, at controlled pressure gradient, would test the rate directly and measure the offset $B_{\NL,G}^r$ that the present narrow $\eta_s$ range holds only through an extrapolation to $\delta=\ks$.

\section{Discussion and conclusions}\label{sec:discussion}
What the fourteen cases establish robustly is spectral, and it belongs at the front.  The energetic population collapses at $T^o\sim 1$ and its shape is Reynolds-independent at fixed geometry; the integrated variance is consistent with growth in $\eta_s=\ln(\delta/\ks)$ --- a positive trend at the $68\%$ but not the $95\%$ level --- rather than demonstrating it.  Behind that collapse is a simple picture: the rough-wall pressure spectrum is the sum of two overlapping log-normal distributions, each modelling a \emph{separate} contribution to the total energy.  One is roughness-local, from the mean-shear (linear) source, and sits at high frequency on the roughness period $T^s=T\ue/\ks$; the other is the energetic contribution from the nonlinear source, at the outer period $T^o=T\ue/\delta\sim 1$, and it alone grows with Reynolds number.  Splitting the total variance into these two contributions is what lets the shape be pinned tightly while the growth rate is predicted rather than fitted.

Each source scales on its own natural length.  The linear source $\Lin$ is roughness-local and contributes a non-growing $O(1)$ offset; the nonlinear source $\NL$ acts over a logarithmic wall-normal range cut off below by the roughness sublayer ($\sim \ks$) and above by $\delta$, so its area carries all the Reynolds-number growth through the single term $\eta_s$, with no $\ks^+$ dependence at leading order.  The fitted outer break $T_{b\NL}^{o,r}\approx 0.98$, with the measured $U_c\approx 0.49\,\ue$ \citep{totten2026}, gives a convected wavelength $\lambda_x\approx 0.5\,\delta$, matching the smooth-wall outer population of \citet{massey_eddy_2025}; kernel consistency favours a log-layer source ($y_{\NL,\mathrm{eff}}/\delta\lesssim 0.1$) over a mid-layer one, whose contribution the elliptic weight $e^{-4\pi}$ would suppress.

The construction makes four predictions, each with a clear experimental test; the offset carries the calibration uncertainty of table~\ref{tab:params}.
\begin{enumerate}
\item \emph{The growth rate is universal.}  Once fully rough, the inner-scaled variance grows on $\eta_s$ at the smooth-wall rate $A_{\NL,\eta}^{\mathrm{base}}\approx 2.3$, with no free parameter: every canopy effect is confined to the non-growing offset $B_{\NL,G}^r$, because anything set by $\ks$ decays in $y_d/\ks$ and integrates over $\dd\ln y_d$ to a finite constant (\S\ref{sec:nonlinear}).  \textit{Test:} a $\delta^+$ sweep at fixed $\ks^+$ over decades, at controlled pressure gradient, measures the slope directly.  A rate significantly above $2.3$ would falsify outer-layer similarity for the pressure source, not calibrate a gain.  On the VT data the free slope is $3.8$ with $68\%$ interval $[1.5,\,6.1]$, consistent with the prediction and unable to discriminate it.
\item \emph{$\ks^+$ drops out once fully rough.}  At fixed geometry all growth is carried by $\delta/\ks$, not by the roughness Reynolds number.  \textit{Test:} vary $\ks^+$ (change roughness size or speed) at matched $\delta/\ks$ and look for a change in the spectrum; there should be none.  The two VT speed families are the only present instance and are inconclusive.
\item \emph{The energetic break is fixed at $T^o\sim 1$.}  It does not move with $\delta/\ks$; its apparent drift on the roughness axis is only the rescaling $T_{b\NL}^{s,r}=(\delta/\ks)\,T_{b\NL}^{o,r}$.  \textit{Test:} track the peak period across a $\delta/\ks$ range --- it should stay fixed in outer units and move in exact proportion to $\delta/\ks$ in roughness units.  The VT collapse at $T^o\sim 1$ already supports this.
\item \emph{Across geometries the growth lines are parallel.}  Geometry enters $B_{\NL,G}^r$ and not the slope, a falsifiable consequence of confining canopy physics to the offset.  \textit{Test:} plot $\overline{p^{\prime 2}}^+$ against $\eta_s$ for different roughness types; the lines should be offset but not tilted.  Existing rough-wall campaigns \citep{blake1970,meyers2015} may already permit this.
\end{enumerate}

These predictions rest on fourteen cases of a single geometry over a short range, and the honest limits are set out in \S\ref{sec:vt}: over $\Delta\eta_s\approx 0.84$ the free-slope test cannot discriminate the predicted rate from twice or half it, and the offset is held only through an extrapolation to $\delta=\ks$; the speed-family probe of the basis is inconclusive, with $\ks^+$ sensitivity \citep{busse2017}, non-equilibrium development \citep{vishwanathan2023} and experimental scatter not separable from the data alone; and with one roughness ($\ks=3.3$~mm cylinders) the use of $\ks$ as the length scale and any across-geometry universality are assumed, not tested.  The correspondence with the geometric-ratio empirical model of \citet{fritsch2023} is accordingly one of spirit, not of cross-geometry validation.

The value of the framework is that the growth rate is predicted outright, at the smooth-wall value on the rough-wall range, and the whole geometry dependence is reduced to one measurable offset.  Three experiments would settle it: the $\delta^+$ sweep at fixed $\ks^+$ (predictions~1 and~3), the $\ks^+$ sweep at matched $\delta/\ks$ (prediction~2), and a comparison across roughness types at matched $\eta_s$ (prediction~4).  The finer decomposition of the offset into sublayer source and canopy Green term needs the wall-normal source statistics that only rough-wall DNS can supply \citep{ma2021,bhaganagar2007}.  In plain terms: roughness fixes the shape of the wall-pressure spectrum, and how fast the fluctuations grow as the boundary layer thickens is governed by how many decades separate the roughness height from the layer thickness.

\backsection[Funding]{The support from DARPA under award HR0011-24-9-0465  is gratefully acknowledged.}
\backsection[Declaration of interests]{The authors report no conflict of interest.}
\backsection[Data availability statement]{The VT data are publicly available on line at: \url{https://data.lib.vt.edu/articles/dataset/Surface_Pressure_Spectra_Beneath_High_Reynolds_Number_Smooth_and_Rough_Wall_Boundary_Layers_in_Pressure_Gradients/20457189}}
\backsection[Author ORCIDs]{J.M.O. Massey, https://orcid.org/0000-0002-2893-955X; A.J. Smits, https://orcid.org/0000-0002-3883-8648; B.J. McKeon, https://orcid.org/0000-0003-4220-1583}

\bibliographystyle{plainnat}
\bibliography{refs}

\end{document}

%% file: figs/model_schematic.tex
\begin{figure}
\centering
\resizebox{\linewidth}{!}{
\begin{tikzpicture}[
    font=\small,
    >={Stealth[length=2.2mm]},
    layer/.style={draw=black!55, line width=0.4pt},
    popbox/.style={rounded corners=2pt, draw=black!70, line width=0.7pt,
                   inner sep=5pt, align=center},
    lab/.style={align=center, font=\footnotesize},
    dep/.style={->, draw=black!60, line width=0.6pt},
    flow/.style={->, draw=black!85, line width=0.9pt},
]
\colorlet{Lcol}{red!72!black}
\colorlet{Qcol}{violet!85!black}

\begin{scope}[xshift=0cm]
  \def\W{3.2}      
  \def\H{5.2}      
  \def\hs{1.35}    
  \def\dh{0.50}    
  \draw[layer] (0,0) rectangle (\W,\H);
  \draw[layer, dashed] (0,\H) -- (\W,\H);
  \node[anchor=east, font=\footnotesize] at (-0.08,\H) {$y_d=\delta$};
  \draw[layer, dashed] (0,\hs) -- (\W,\hs);
  \node[anchor=east, font=\footnotesize, align=right] at (-0.08,\hs)
      {$y_d\!\approx\!3\ks$};
  \draw[layer, dotted] (0,\dh) -- (\W,\dh);
  \node[anchor=east, font=\footnotesize] at (-0.08,\dh) {$y_d\!=\!0$};
  \draw[<->, black!70, line width=0.5pt] (0.13,0.02) -- (0.13,\dh-0.02);
  \node[anchor=west, font=\scriptsize, black!70] at (0.15,\dh/2) {$d$};
  \foreach \x in {0.42,1.02,1.62,2.22,2.82}{
    \draw[fill=black!12, draw=black!55, line width=0.4pt]
        (\x,0) rectangle ++(0.24,0.70);
  }
  \node[anchor=east, font=\footnotesize] at (-0.08,0.31) {$\ks$};
  \draw[black!55, dashed, line width=1.0pt]
      plot[smooth, domain=0.95:\H, samples=40]
      ({0.20 + 1.15*ln((\x)/0.62) + 0.72*(1 - (\x/\H)*(\x/\H))}, \x);
  \draw[black!88, line width=1.3pt]
      plot[smooth, domain=0.72:\H, samples=40]
      ({0.20 + 1.15*ln((\x)/0.62)}, \x);
  \node[anchor=west, font=\footnotesize, black!88] at (2.62,\H-0.16) {$\ue$};
  \draw[<->, black!78, line width=0.8pt] (1.83,2.55) -- (2.37,2.55);
  \node[anchor=south, font=\footnotesize, black!78] at (2.02,2.10) {$\Delta U^+$};
  \draw[black!88, line width=1.3pt] (0.16,4.78) -- (0.60,4.78);
  \node[anchor=west, font=\scriptsize, black!88] at (0.63,4.78) {rough};
  \draw[black!55, dashed, line width=1.0pt] (0.16,4.40) -- (0.60,4.40);
  \node[anchor=west, font=\scriptsize, black!55] at (0.63,4.40) {smooth};

  \node[lab, Lcol, anchor=south east] at (\W-0.08,0.12)
      {roughness\\[-1pt]sublayer};
  \node[lab, Qcol, anchor=west, text width=1.05cm] at (0.05,3.55)
      {inertial\\[-1pt] region};
  \node[font=\bfseries\footnotesize, anchor=south] at (\W/2,-0.95)
      {(a) mean velocity};
  \node[font=\footnotesize, anchor=south] at (\W/2,-1.35)
      {and scales};
\end{scope}

\begin{scope}[xshift=6.5cm]
  \node[popbox, draw=Qcol, fill=Qcol!8, text width=3.7cm] (Q) at (0.5,3.7)
    {\textbf{Nonlinear} $\NL$\\[1pt] inertial-layer source};
  \node[popbox, draw=Lcol, fill=Lcol!8, text width=3.7cm] (L) at (0.5,1.0)
    {\textbf{Linear} $\Lin$\\[1pt] roughness-local};
  \node[font=\bfseries\footnotesize, anchor=south] at (0,-0.95)
      {(b) source};
  \node[font=\footnotesize, anchor=south] at (0,-1.35)
      {populations};
\end{scope}
\draw[flow, draw=Qcol] (3.25,3.2) to[out=15,in=180] (Q.west);
\draw[flow, draw=Lcol] (3.25,0.7) to[out=-5,in=180] (L.west);

\begin{scope}[xshift=10.6cm]
  \def\AX{3.7}\def\AY{2.0}
  \draw[->, black!70, line width=0.6pt] (0,0) -- (\AX+0.2,0)
     node[anchor=west, font=\footnotesize] {$\log T$};
  \draw[->, black!70, line width=0.6pt] (0,0) -- (0,\AY+0.3)
     node[anchor=south, font=\footnotesize] {${f\phi_{pp}}^+$};
  \draw[Lcol, line width=1.0pt, fill=Lcol!12]
     plot[smooth, domain=0.15:1.7, samples=40]
     ({\x}, {0.55*exp(-((\x-0.9)^2)/(2*0.16))});
  \draw[Qcol, line width=1.0pt, fill=Qcol!12]
     plot[smooth, domain=1.5:3.6, samples=40]
     ({\x}, {1.75*exp(-((\x-2.6)^2)/(2*0.28))});
  \node[font=\footnotesize, anchor=south, Lcol] at (0.9,0.62) {$g_{\Lin}^r$};
  \node[font=\footnotesize, anchor=south, Qcol] at (2.6,1.80) {$g_{\NL}^r$};
  \node[anchor=north, font=\scriptsize, Lcol] at (0.75,-0.12)
      {$T^s\!=\!T\ue/\ks$};
  \node[anchor=north, font=\scriptsize, Qcol] at (2.85,-0.12)
      {$T^o\!=\!T\ue/\delta$};
  \node[font=\bfseries\footnotesize, anchor=south] at (\AX/2,-1.55)
      {(c) spectral};
  \node[font=\footnotesize, anchor=south] at (\AX/2,-1.95)
      {placement};
\end{scope}

\begin{scope}[yshift=-2.55cm]
  \node[popbox, draw=black!45, fill=none, text width=4.5cm, anchor=north]
       (grow) at (5.0,0)
    {\footnotesize \textbf{Re-growth carrier}\\[2pt]
     nonlinear area $\propto \eta_s=\ln(\delta/\ks)$};
  \node[popbox, draw=black!45, fill=none, text width=4.3cm, anchor=north]
       (off) at (10.6,0)
    {\footnotesize \textbf{non-growing offsets}\\[2pt]
     $B_{\Lin}^r,\;\; B_{\NL,G}^r$};
\end{scope}
\draw[dep, draw=Qcol] ([yshift=0.2cm]Q.south west) -| ([xshift=-0.8cm]grow.north);
\draw[dep, draw=Qcol] (Q.south east) .. controls +(1.7,-0.7) and +(0,1.7) .. ([xshift=-0.2cm]off.north);
\draw[dep, draw=Lcol] (L.east) -| ([xshift=-0.7cm]off.north);

\end{tikzpicture}
}
\caption{Two-population structure of the fully rough wall-pressure spectrum,
with the mean velocity and length scales used in the text.
(a)~Mean velocity and scales.  The wall-normal coordinate $y_d=y-d$ is measured
from the displacement plane $y_d=0$, a height $d$ (the displacement height)
above the true wall.  Above the roughness sublayer the rough mean profile
(solid) is the smooth-wall logarithmic profile (dashed) shifted to lower
velocity by the roughness function $\Delta U^+$; the two merge to the edge
velocity $\ue$ at $y_d=\delta$.  The length scales are the roughness height
$\ks$, the layer thickness $\delta$, and the sublayer cutoff
$(\sim 3\ks)$.  The roughness sublayer ($y_d\lesssim 3\ks$) is the seat
of the roughness-local linear source $\Lin$ (red); the inertial region above it
is the seat of the nonlinear source $\NL$ (purple).  (b)~The two source
populations.  (c)~They map to distinct premultiplied-spectrum peaks: the
roughness-local $g_{\Lin}^r$ (red), fixed on the roughness period
$T^s=T\ue/\ks$, and the energetic $g_{\NL}^r$ (purple), on the outer period
$T^o=T\ue/\delta$.}
\label{fig:schematic}
\end{figure}

%% file: refs.bib
@article{kim1989,
  author  = {Kim, John},
  title   = {On the structure of pressure fluctuations in simulated turbulent channel flow},
  journal = {Journal of Fluid Mechanics},
  volume  = {205},
  pages   = {421--451},
  year    = {1989},
  doi     = {10.1017/S0022112089002090}
}

@article{changpiomelli1999,
  author  = {Chang, Peter A. and Piomelli, Ugo and Blake, William K.},
  title   = {Relationship between wall pressure and velocity-field sources},
  journal = {Physics of Fluids},
  volume  = {11},
  number  = {11},
  pages   = {3434--3448},
  year    = {1999},
  doi     = {10.1063/1.870202}
}

@article{anantharamu2020,
  author  = {Anantharamu, Sreevatsa and Mahesh, Krishnan},
  title   = {Analysis of wall-pressure fluctuation sources from direct numerical
             simulation of turbulent channel flow},
  journal = {Journal of Fluid Mechanics},
  volume  = {898},
  pages   = {A17},
  year    = {2020},
  doi     = {10.1017/jfm.2020.412}
}

@article{panton2017,
  author  = {Panton, Ronald L. and Lee, Myoungkyu and Moser, Robert D.},
  title   = {Correlation of pressure fluctuations in turbulent wall layers},
  journal = {Physical Review Fluids},
  volume  = {2},
  number  = {9},
  pages   = {094604},
  year    = {2017},
  doi     = {10.1103/PhysRevFluids.2.094604}
}

@article{schlatterorlu2010,
  author  = {Schlatter, Philipp and {\"O}rl{\"u}, Ramis},
  title   = {Assessment of direct numerical simulation data of turbulent boundary layers},
  journal = {Journal of Fluid Mechanics},
  volume  = {659},
  pages   = {116--126},
  year    = {2010},
  doi     = {10.1017/S0022112010003113}
}

@article{leemoser2015,
  author  = {Lee, Myoungkyu and Moser, Robert D.},
  title   = {Direct numerical simulation of turbulent channel flow up to
             $\mathrm{Re}_\tau \approx 5200$},
  journal = {Journal of Fluid Mechanics},
  volume  = {774},
  pages   = {395--415},
  year    = {2015},
  doi     = {10.1017/jfm.2015.268}
}

@article{massey2026nl,
  author  = {Massey, Jonathan M. O. and Klewicki, Joseph C. and McKeon, Beverley J.},
  title   = {On the {P}oisson-source basis of logarithmic wall-pressure-variance growth},
  journal = {arXiv preprint arXiv:2511.16776},
  year    = {2026}
}

@article{massey2026inner,
  title={An Inner-Scaled Linear Contribution to Wall-Pressure Variance at High Reynolds Number},
  author={Massey, JMO and Zimmerman, SJ and Klewicki, JC and McKeon, BJ},
  journal={arXiv preprint arXiv:2607.02395},
  year={2026}
}

@article{bradshaw1967,
  author  = {Bradshaw, P.},
  title   = {`{I}nactive' motion and pressure fluctuations in turbulent boundary layers},
  journal = {Journal of Fluid Mechanics},
  volume  = {30},
  number  = {2},
  pages   = {241--258},
  year    = {1967},
  doi     = {10.1017/S0022112067001417}
}

@article{pantonlinebarger1974,
  author  = {Panton, Ronald L. and Linebarger, John H.},
  title   = {Wall pressure spectra calculations for equilibrium boundary layers},
  journal = {Journal of Fluid Mechanics},
  volume  = {65},
  number  = {2},
  pages   = {261--287},
  year    = {1974},
  doi     = {10.1017/S0022112074001388}
}

@article{jimenez2004,
  author  = {Jim{\'e}nez, Javier},
  title   = {Turbulent flows over rough walls},
  journal = {Annual Review of Fluid Mechanics},
  volume  = {36},
  pages   = {173--196},
  year    = {2004},
  doi     = {10.1146/annurev.fluid.36.050802.122103}
}

@book{townsend1976,
  author    = {Townsend, A. A.},
  title     = {The Structure of Turbulent Shear Flow},
  edition   = {2nd},
  publisher = {Cambridge University Press},
  address   = {Cambridge},
  year      = {1976}
}

@article{hama1954,
  author  = {Hama, F. R.},
  title   = {Boundary-layer characteristics for smooth and rough surfaces},
  journal = {Transactions of the Society of Naval Architects and Marine Engineers},
  volume  = {62},
  pages   = {333--358},
  year    = {1954}
}

@article{massey_eddy_2025,
  author  = {Massey, Jonathan M. O. and Smits, Alexander J. and McKeon, Beverley J.},
  title   = {Two-component inner--outer scaling model for the wall-pressure spectrum
             at high {R}eynolds number},
  journal = {Journal of Fluid Mechanics},
  volume  = {1034},
  pages   = {A60},
  year    = {2026},
  doi     = {10.1017/jfm.2026.11550}
}

@article{schultz2007,
  author  = {Schultz, M. P. and Flack, K. A.},
  title   = {The rough-wall turbulent boundary layer from the hydraulically smooth
             to the fully rough regime},
  journal = {Journal of Fluid Mechanics},
  volume  = {580},
  pages   = {381--405},
  year    = {2007},
  doi     = {10.1017/S0022112007005502}
}

@article{mehdi2013,
  author  = {Mehdi, Faraz and Klewicki, Joseph C. and White, Christopher M.},
  title   = {Mean force structure and its scaling in rough-wall turbulent boundary layers},
  journal = {Journal of Fluid Mechanics},
  volume  = {731},
  pages   = {682--712},
  year    = {2013},
  doi     = {10.1017/jfm.2013.385}
}

@article{raupach1991,
  author  = {Raupach, M. R. and Antonia, R. A. and Rajagopalan, S.},
  title   = {Rough-wall turbulent boundary layers},
  journal = {Applied Mechanics Reviews},
  volume  = {44},
  number  = {1},
  pages   = {1--25},
  year    = {1991},
  doi     = {10.1115/1.3119492}
}

@article{raupach1982,
  author  = {Raupach, M. R. and Shaw, R. H.},
  title   = {Averaging procedures for flow within vegetation canopies},
  journal = {Boundary-Layer Meteorology},
  volume  = {22},
  number  = {1},
  pages   = {79--90},
  year    = {1982},
  doi     = {10.1007/BF00128057}
}

@article{squire2016,
  author  = {Squire, D. T. and Morrill-Winter, C. and Hutchins, N. and
             Marusic, I. and Schultz, M. P. and Klewicki, J. C.},
  title   = {Smooth- and rough-wall boundary layer structure from high spatial range
             particle image velocimetry},
  journal = {Physical Review Fluids},
  volume  = {1},
  number  = {6},
  pages   = {064402},
  year    = {2016},
  doi     = {10.1103/PhysRevFluids.1.064402}
}

@article{busse2017,
  author  = {Busse, A. and Thakkar, M. and Sandham, N. D.},
  title   = {Reynolds-number dependence of the near-wall flow over irregular rough surfaces},
  journal = {Journal of Fluid Mechanics},
  volume  = {810},
  pages   = {196--224},
  year    = {2017},
  doi     = {10.1017/jfm.2016.680}
}

@article{aghaeijouybari2022,
  author  = {Aghaei-Jouybari, Mostafa and Seo, Jung-Hee and Yuan, Junlin and
             Mittal, Rajat and Meneveau, Charles},
  title   = {Contributions to pressure drag in rough-wall turbulent flows:
             Insights from force partitioning},
  journal = {Physical Review Fluids},
  volume  = {7},
  number  = {8},
  pages   = {084602},
  year    = {2022},
  doi     = {10.1103/PhysRevFluids.7.084602}
}

@article{zhang2026,
  author  = {Zhang, Zichun and Feng, Zexin and Huo, Kebing and Jiang, Nan},
  title   = {Outer-layer similarity from the perspective of uniform momentum zones
             in turbulent boundary layer over smooth and rough wall},
  journal = {International Journal of Heat and Fluid Flow},
  volume  = {119},
  pages   = {110292},
  year    = {2026},
  doi     = {10.1016/j.ijheatfluidflow.2026.110292}
}

@article{meyers2015,
  author  = {Meyers, Timothy and Forest, Jonathan B. and Devenport, William J.},
  title   = {The wall-pressure spectrum of high-{R}eynolds-number turbulent
             boundary-layer flows over rough surfaces},
  journal = {Journal of Fluid Mechanics},
  volume  = {768},
  pages   = {261--293},
  year    = {2015},
  doi     = {10.1017/jfm.2014.743}
}

@inproceedings{catlett2022,
  author    = {Catlett, Matthew R. and Bryan, Benjamin S. and Chang, Natasha
               and Hemingway, Hugh and Anderson, Jason M.},
  title     = {Modeling unsteady surface pressure auto-spectra for turbulent
               boundary layer flow over small dense roughness},
  booktitle = {AIAA SCITECH 2022 Forum},
  year      = {2022},
  doi       = {10.2514/6.2022-2560},
  note      = {AIAA Paper 2022-2560}
}

@article{farabee1991,
  author  = {Farabee, Theodore M. and Casarella, Mario J.},
  title   = {Spectral features of wall pressure fluctuations beneath turbulent
             boundary layers},
  journal = {Physics of Fluids A: Fluid Dynamics},
  volume  = {3},
  number  = {10},
  pages   = {2410--2420},
  year    = {1991},
  doi     = {10.1063/1.858179}
}

@article{bull1996,
  author  = {Bull, M. K.},
  title   = {Wall-pressure fluctuations beneath turbulent boundary layers:
             some reflections on forty years of research},
  journal = {Journal of Sound and Vibration},
  volume  = {190},
  number  = {3},
  pages   = {299--315},
  year    = {1996},
  doi     = {10.1006/jsvi.1996.0066}
}

@article{joseph2022,
  author  = {Joseph, Liselle A. and Devenport, William J. and Glegg, Stewart},
  title   = {Empirical model for low-speed rough-wall turbulent boundary layer
             pressure spectra},
  journal = {AIAA Journal},
  volume  = {60},
  number  = {4},
  pages   = {2160--2168},
  year    = {2022},
  doi     = {10.2514/1.J060965}
}

@article{vishwanathan2023,
  author  = {Vishwanathan, Vidya and Fritsch, Daniel J. and Lowe, K. Todd and Devenport, William J.},
  title   = {History effects and wall-similarity of non-equilibrium turbulent boundary layers
             in varying pressure gradient over rough and smooth surfaces},
  journal = {International Journal of Heat and Fluid Flow},
  volume  = {102},
  pages   = {109145},
  year    = {2023},
  doi     = {10.1016/j.ijheatfluidflow.2023.109145}
}

@article{fritsch2023,
  author  = {Fritsch, Daniel J. and Vishwanathan, Vidya and Roy, Christopher J.
             and Lowe, K. Todd and Devenport, William J.},
  title   = {Modeling the surface pressure spectrum on rough walls in pressure gradients},
  journal = {Journal of Fluids Engineering},
  volume  = {145},
  number  = {12},
  pages   = {121301},
  year    = {2023},
  doi     = {10.1115/1.4062821}
}

@article{totten2026,
  author  = {Totten, Eric and Damani, Shishir and Sharma, Bhavika and Butt, Humza
             and Rawther, Celin and Glegg, Stewart and Devenport, William
             and Lowe, Todd},
  title   = {Measurements and model comparison of sub-convective pressure fluctuations
             in rough wall turbulent boundary layers},
  journal = {Journal of Fluid Mechanics},
  volume  = {1036},
  pages   = {A33},
  year    = {2026},
  doi     = {10.1017/jfm.2026.11640}
}

@article{goody2004,
  author  = {Goody, Michael},
  title   = {Empirical spectral model of surface pressure fluctuations},
  journal = {AIAA Journal},
  volume  = {42},
  number  = {9},
  pages   = {1788--1794},
  year    = {2004},
  doi     = {10.2514/1.9433}
}

@article{blake1970,
  author  = {Blake, William K.},
  title   = {Turbulent boundary-layer wall-pressure fluctuations on smooth and rough walls},
  journal = {Journal of Fluid Mechanics},
  volume  = {44},
  number  = {4},
  pages   = {637--660},
  year    = {1970},
  doi     = {10.1017/S0022112070002069},
}

@article{ma2021,
  author  = {Ma, Rong and Alam{\'e}, Karim and Mahesh, Krishnan},
  title   = {Direct numerical simulation of turbulent channel flow over random rough surfaces},
  journal = {Journal of Fluid Mechanics},
  volume  = {908},
  pages   = {A40},
  year    = {2021},
  doi     = {10.1017/jfm.2020.874},
}

@article{bhaganagar2007,
  author  = {Bhaganagar, Kiran and Coleman, Gary N. and Kim, John},
  title   = {Effect of roughness on pressure fluctuations in a turbulent channel flow},
  journal = {Physics of Fluids},
  volume  = {19},
  number  = {2},
  pages   = {028103},
  year    = {2007},
  doi     = {10.1063/1.2482883},
}

@article{howe1988,
  author  = {Howe, M. S.},
  title   = {The turbulent boundary-layer rough-wall pressure spectrum at
             acoustic and subconvective wavenumbers},
  journal = {Proceedings of the Royal Society A},
  volume  = {415},
  number  = {1848},
  pages   = {141--161},
  year    = {1988},
  doi     = {10.1098/rspa.1988.0007}
}

@article{gleggdevenport2009,
  author  = {Glegg, Stewart A. L. and Devenport, William J.},
  title   = {The far-field sound from rough-wall boundary layers},
  journal = {Proceedings of the Royal Society A},
  volume  = {465},
  number  = {2106},
  pages   = {1717--1734},
  year    = {2009},
  doi     = {10.1098/rspa.2008.0318}
}
